\documentclass[seceq,preprint]{ptptex}

\usepackage[dvips]{graphicx}
\usepackage[dvips]{psfrag}
\usepackage{amsmath}
\usepackage{amsfonts}

\notypesetlogo                       

\title{
Holography of Two-point Functions in \\ the Spinning String/Spin Chain Correspondence%
}

\author{
Akihiro \textsc{Tsuji}\thanks{E-mail: tsuji@hep1.c.u-tokyo.ac.jp}%
}

\inst{
Institute of Physics, University of Tokyo, Komaba \\
Tokyo 153-8902, Japan
}

\recdate{October 19, 2006}            

\abst{
We study the holographic principle of the two-point function in the case of a spinning string in the AdS/CFT correspondence. Following the general method proposed by Yoneya et al. for the bulk-boundary correspondence in the large $J$ limit, 
we study the spinning string solution which starts and ends at the boundary. We then show that the spinning string solution directly gives a two-point function that is consistent with the spin chain picture in gauge theory.
}

\begin{document}

\maketitle
\section{Introduction}
One of the most remarkable aspects of the AdS/CFT correspondence \cite{9711200,9802109,9802150}(see Refs. 4) and 5) for reviews) is that it can be regarded as a holographic relation between the bulk string theory and the corresponding gauge theory on the boundary. The GKP-Witten relation \cite{9802109,9802150} gives the explicit connection between the physical quantities involved in particular the correlation functions. Also, it is the power of the AdS/CFT correspondence that this correspondence can be considered a weak/strong duality, and thus we may learn something about strong coupling physics, such as confinement, by using AdS/CFT. For this reason, we are particularly interested in the dynamics of Wilson loops in the bulk picture.
Wilson loops are considered to be the ends of the string worldsheet in the bulk picture, 
and we can determine the vacuum expectation value of the Wilson loops by calculating the area of the worldsheet \cite{9803001,9803002}. In recent, various surfaces corresponding to several Wilson loops have been studied \cite{9809188,9904191,9904149,0205160,0207241,0211229,0604058,Drukker:2006xg}. However, since the AdS/CFT correspondence is effective 
in the region of strong coupling in gauge theory, it is difficult to compare bulk quantities with the boundary values, 
except in highly supersymmetric cases, in which higher-loop effects can be ignored.

Concerning the relation between AdS quantities and CFT quantities, Berenstein, Maldacena and Nastase proposed a method to compare the string energy spectrum with the conformal dimensions of boundary operators directly for the special case \cite{0202021}, in which the string has an infinitely large orbital angular momentum, $J \sim R^2$. They took the limit $J, \lambda \rightarrow \infty$, with $\tilde{\lambda} \equiv \frac{\lambda}{J^2} \ll 1$ fixed. Since the $\tilde{\lambda}$ expansion is effective in both string theory and gauge theory, we can compare physical quantities in the two cases order by order. The corresponding operators in gauge theory are called BMN operators. The BMN operators are still near-BPS. Subsequently, the correspondence was generalized to operators that are far from BPS, having the same large $J$ limit, but with more charges of non-BPS operators \cite{0212208,0307042}. This is known as the spinning string/spin chain correspondence \cite{0308117}.

As a further development of the spinning string/spin chain correspondence, it has been found that the string nonlinear sigma model action is identical to the continuum coherent state effective action of a spin chain \cite{0311203,0403120}. The mapping of a spinning string/spin chain is derived by directly relating each coordinate with a spin chain field. Specifically, the string configuration corresponds to the coherent state expectation value of the spin chain, namely, the `long' scalar local operator. 
Thus, the action of the string non-linear sigma model coincides with the coherent state effective action of the spin chain.

In these approaches, the string is always at the center of AdS space. However, from the viewpoint of holography, these relations should be realized as the correspondence between the bulk of AdS space and its conformal boundary, as formulated by the GKP-Witten relation for correlation functions. In the case of the BMN limit, this puzzle was resolved by Dobashi, Shimada and Yoneya \cite{Dobashi:2002ar,Yoneya:2003mu} by employing a tunneling picture, in which the correlators are directly obtained by considering the bulk trajectories from the boundary to the boundary. Using this method, they proposed a definite holographic relation for 3-point correlation functions \cite{Dobashi:2004nm,Dobashi:2004ka,Dobashi:2006fu,Shimada:2004sw}.

In the present paper, we investigate the holographic two-point functions which are related to the spinning strings. The holography of the one-point function of a single Wilson loop is studied in Refs. 6) and 7), in which the boundary of the string worldsheet is regarded as the Wilson loop, and the one-point function is obtained by calculating the area of the worldseet. In our case, the boundary operator cannot be regarded as the usual Wilson loop, because it has fixed R-charges. We can assume that the states corresponding to the boundaries of the worldseet are the spin chain coherent states from Refs. 21) and 22). We can then compute the two-point function of the boundary operator by calculating the action of the `tunneling' spinning string solution. This approach is useful for clarifying the holographic interpretation of spinning strings.

In $\S$2, we study the tunneling picture of the spinning string solution after reviewing the spinning string solution with $S^5$ angular momenta $(J_1,J_2)$ and the method of Yoneya et al. In $\S$3, we calculate the action of the tunneling spinning string solution and obtain the two-point correlation function of the boundary operator in the holographic description. We find that this boundary operator has a definite conformal dimension. This result has a very natural interpretation from the viewpoint of the configuration mapping discussed in Refs. 21) and 22). We regard the radial direction of AdS space as the affine time direction of the Euclideanized geodesic, while the endpoint at the boundary corresponds to spin chain coherent states. 
The spin chain states have definite conformal dimensions, once the appropriate Bethe root configuration is given. Finally, we conclude our work and mention some future directions in $\S$4.

\section{`Tunneling' spinning string solution}
\subsection{Spinning string solution}
First, we review the usual spinning string solution with two large angular momenta $J_1$ and $J_2$ \cite{0304255,0306130}.

The Polyakov action is
\begin{equation}
S=\frac{\sqrt{\lambda}}{2}\int d \tau \int \frac{d \sigma}{2 \pi} \sqrt{-g}G^{\mu \nu}g^{ab}\partial_{a}X_{\mu}\partial_b X_{\nu},
\end{equation}
where $g$ is the worldsheet metric and $G$ is the space-time metric.

The $AdS_5 \times S^5$metric is $ds^2=(ds^2)_{AdS_5} + (ds^2)_{S^5}$ where
\begin{eqnarray}
(ds^2)_{AdS_5} &=& d \rho ^2-\cosh^2{\rho}dt_g^2+\sinh^2{\rho}(d \theta^2+\cos^2{\theta}d \phi_1^2+\sin^2{\theta}d \phi^2_2), \nonumber \\
(ds^2)_{S^5} &=& d \gamma^2+\cos^2{\gamma}d \varphi^2_3+\sin^2{\gamma}(d \psi^2+\cos^2{\psi}d \varphi^2_1+\sin^2{\psi}d \varphi^2_2)
\end{eqnarray}
in global coordinates. Note that we denote the golobal time coordinate by $t_g$. We choose the gauge such that $\det g_{ab}=-1, g_{01}=0, g^{00}=-\eta, \partial_{\sigma}\eta=0$. Then, we choose the solution of the equation of motion as 
\begin{eqnarray}
t_g &=& \kappa \tau, \phi_1=\omega_1\tau, \phi_2=\omega_2\tau, \nonumber \\
\varphi_1 &=& w_1\tau, \varphi_2=w_2\tau, \varphi_3=w_3\tau. 
\end{eqnarray}
The other variables, $\rho, \theta, \gamma$ and $\psi$ depend only on the $\sigma$ coordinate and satisfy the boundary conditions of a closed string.

With the system described above, we have five conserved charges and the charge of the global time translation given by 
\begin{eqnarray}
S_1=\sqrt{\lambda} \eta \omega_1\int^{2\pi}_0\frac{d \sigma}{2\pi}\sinh^2{\rho}\cos^2{\theta}, \hspace{2pt}
&&J_1=\sqrt{\lambda} \eta w_1\int^{2\pi}_0\frac{d \sigma}{2\pi}\sin^2{\gamma}\cos^2{\psi}, \nonumber \\
S_2=\sqrt{\lambda} \eta \omega_2\int^{2\pi}_0\frac{d \sigma}{2\pi}\sinh^2{\rho}\sin^2{\theta}, \hspace{2pt}
&&J_2=\sqrt{\lambda} \eta w_2\int^{2\pi}_0\frac{d \sigma}{2\pi}\sin^2{\gamma}\sin^2{\psi}, \nonumber \\
E=\sqrt{\lambda} \eta \kappa\int^{2\pi}_0\frac{d \sigma}{2\pi}\cosh^2{\rho}, \hspace{32pt}
&&J_3=\sqrt{\lambda} \eta w_3\int^{2\pi}_0\frac{d \sigma}{2\pi}\cos^2{\gamma}, 
\end{eqnarray}
where $\lambda$ is the 't Hooft coupling constant. From the equation of motion, we can choose 
\begin{eqnarray}
\rho &=& \theta=0, \nonumber \\
\gamma &=& \frac{\pi}{2}. 
\end{eqnarray}
This solution possesses two nonzero angular momenta. The equation of motion for $\psi$ then becomes
\begin{equation}
\psi ''-\eta^2(w_1^2-w_2^2)\sin{\psi}\cos{\psi}=0. \label{epsia}
\end{equation}
Here the primes represents differentiation with respect to $\sigma$.

From the equation of motion for $\eta$, we get 
\begin{equation}
-\kappa^2+\frac{{\psi '}^2}{\eta^2}+w_1^2\cos^2{\psi}+w_2^2\sin^2{\psi}=0. \label{viraa}
\end{equation}
This equation connects the $AdS$ parameter to the $S^5$ parameters. Of course, this equation is consistent with the gauge fixing conditions that $\eta$ be independent of  $\sigma$, as seen from (\ref{epsia}). Also, it is clear from the form of (\ref{viraa}) that $\eta$ does not depend on $\tau$.

We solve these equations in the case of a folded string. 
We assume
\begin{equation}
0 \le \frac{\kappa^2-w_1^2}{w_2^2-w_1^2}=\sin^2{\psi_0}\equiv x \le 1,
\end{equation}
where $\psi_0$ is the value of $\psi$ at which $\psi ' =0$. Then we can compute
\begin{eqnarray}
\frac{J_1}{\sqrt{\lambda}}&=&w_1\int_0^{2\pi}\frac{d \sigma}{2 \pi}\eta \cos^2{\psi}\nonumber \\
&=&\frac{2w_1}{\pi}\int_0^{\psi_0}d \psi \frac{\eta \cos^2{\psi}}{\psi '}\nonumber \\
&=&\frac{2w_1}{\pi \sqrt{w_2^2-w_1^2}}\int_0^{\psi_0}d \psi \frac{\cos^2{\psi}}{\sqrt{x-\sin^2{\psi}}}\nonumber \\
&=&\frac{w_1}{\sqrt{w_2^2-w_1^2}}{}_2F_1 \left(-\frac{1}{2}, \frac{1}{2}, 1;x \right)=\frac{2w_1}{\pi \sqrt{w_2^2-w_1^2}}E(x),
\end{eqnarray}
where ${}_2F_1(\alpha, \beta, \gamma;z)$ is the hypergeometric function and $E(x)$ is the elliptic integral of the second kind. Similarly, we have
\begin{equation}
\frac{J_2}{\sqrt{\lambda}}=\frac{w_2x}{2 \sqrt{w_2^2-w_1^2}}{}_2F_1 \left(\frac{1}{2}, \frac{3}{2}, 2;x \right), 
\end{equation}
and $E= \sqrt{\lambda} \eta \kappa$.

Using these equations and $J_1+J_2=J$, we can write $x$ and  $E$ in terms of $J_1$ and $J_2$. 
In order to do so, we must solve the equations
\begin{equation}
\left(\frac{E}{K(x)}\right)^2-\left(\frac{J_1}{E(x)}\right)^2=\frac{4 \lambda}{\pi^2}x, 
\left(\frac{J_2}{K(x)-E(x)}\right)^2-\left(\frac{J_1}{E(x)}\right)^2=\frac{4 \lambda}{\pi^2}. 
\end{equation}
Here, $K(x)$ is the elliptic integral of the first kind. Now, we take the BMN limit $(\lambda,J \rightarrow \infty,\frac{\lambda}{J^2} \equiv \tilde{\lambda} \ll 1)$ and expand
\begin{eqnarray} 
x &=& x_0+ \frac {x_1}{\mathcal{J}^2}+ \frac{x_2}{\mathcal{J}^4 }+... , \\
E &=& J(1+\frac{\lambda}{J^2}\epsilon_1+\frac{\lambda^2}{J^4}\epsilon_2+...)+\mathcal{O}(J^0)\label{eq:eex}. 
\end{eqnarray}
Then we solve the equations order by order and get
\begin{eqnarray}
\frac{E(x_0)}{K(x_0)} &=& \frac{J_1}{J}, \label{as}\\
x_1 &=& -\frac{4(1-x_0)x_0(K(x_0)-E(x_0))E(x_0)K(x_0)^2}{\pi^2((1-x_0)K(x_0)^2-2(1-x_0)K(x_0)E(x_0)+E(x_0)^2)}, \\
\epsilon_1 &=& \frac{2}{\pi^2}K(x_0)(E(x_0)-(1-x_0)K(x_0))\label{es}. 
\end{eqnarray}
Here, $\epsilon_1$ and $\epsilon_2$ are known to coincide with the one- and two-loop $\tilde{\lambda}$ order expansion of the spin chain energy, respectively, calculated by using the Bethe ansatz. It is known that the three-loop expansion is inconsistent, and this is called `the three-loop discrepancy'. Explaining these mismatches remains an important unsolved problem.

\subsection{Tunneling picture of a spinning string}
In this solution, the string is at the center of AdS space, but we want a solution which comes from the boundary and goes to the boundary from the holographic point of view. As usual, such a solution does not exist, because there is a potential barrier near the boundary. In quantum mechanics, a solution that goes through a potential barrier represents a tunnelling effect, and we can obtain such a solution by applying a Wick rotation to the time coordinate. It is shown in Ref. 25) that this tunneling picture \cite{Dobashi:2002ar,Yoneya:2003mu} automatically arises if we directly study the large $J$ limit of the GKP-Witten relation. 
Similarly, we can derive a solution for a spinning string from the boundary to the boundary by applying Wick rotations to the worldsheet $\tau$ coordinate and the $t$ coordinate in target space, where $t$ is the boundary time variable in Poincar\`e coordinates. Also, we apply Wick rotations to the $\varphi_1$ and $\varphi_2$ coordinates in target space to keep the solution a real function. As a result, the quantum numbers are kept real.

The $AdS_5$ metric in Poincar\`e coordinates is
\begin{equation}
ds^2=\frac{dz^2+dx^idx_i-dt^2}{z^2}, 
\end{equation}
and the $S^5$ metric is that used above. Since we consider a spinning string, we assume that it is point-like in AdS space and has extension only in the $S^5$ directions. That is, we assume that $z$ and $t$ depend only on the $\tau$ coordinate. 
Also, we assume
\begin{equation}
\varphi_1=w_1 \tau, \varphi_2=w_2 \tau
\end{equation}
for the center of mass. As in the previous subsection, $\psi$ and $\theta$ depend only on the $\sigma$ coordinate. Then we perform
 Wick rotations as $\tau \rightarrow -i \tau, t \rightarrow -it, 
\varphi_1 \rightarrow -i \varphi_1, \varphi_2 \rightarrow -i \varphi_2$.

Employing the same gauge as in the previous subsection, the equations of motion are
\begin{eqnarray}
 -\frac{\dot{z}^2+\dot{t}^2}{z^2}+\frac{{\psi '}^2}{\eta^2}+w_1^2\cos^2{\psi}+w_2^2\sin^2{\psi}&=&0, \label{etaeom}\\
\frac{\eta}{z^3} \left(\dot{z}^2+\dot{t}^2+z \ddot{z}-2 \dot{z}^2-\frac{\dot{\eta}}{\eta}z{\dot{z}} \right)&=&0, \\
\frac{\partial}{\partial \tau}\left( \eta \frac{\dot{t}}{z^2}\right)&=&0, 
\label{tuneom}\\
\psi ''-\eta^2 (w_1^2-w_2^2)\sin{\psi}\cos{\psi}&=&0 \label{eompsi}, 
\end{eqnarray}
where we take the spinning string solution for the $S^5$ coordinates, that is
\begin{equation}
\gamma=\frac{\pi}{2}, \theta=\varphi_3=0.
\end{equation}
A solution which comes from the boundary and goes to the boundary is
\begin{eqnarray}
t&=&\alpha \tanh{\beta \tau},\\
z&=&\frac{\alpha}{\cosh{\beta \tau}}, \label{soltun}
\end{eqnarray}
where $\alpha$ and $\beta$ are unfixed parameters. In this solution, $\eta$ does not depend on $\tau$ as seen from Eq. (\ref{etaeom}).

Substituting the solution (\ref{soltun}) into (\ref{etaeom}), we get
\begin{eqnarray}
&& -\frac{\dot{z}^2+\dot{t}^2}{z^2}+\frac{{\psi '}^2}{\eta^2}+w_1^2\cos^2{\psi}+w_2^2\sin^2{\psi} \nonumber\\
&& =- \beta^2+\frac{{\psi '}^2}{\eta^2}+w_1^2\cos^2{\psi}+w_2^2\sin^2{\psi}=0.  \label{vira}
\end{eqnarray}
Equation (\ref{eompsi}) and (\ref{vira}) are identical to the usual folded string solution, given by (\ref{epsia}) and (\ref{viraa}), if we replace $\beta$ by the global time translation parameter $\kappa$. Hence, we can calculate $\beta$ using the same method as in the previous subsection, by solving for $\kappa$ in the usual Minkowski picture. 
Because $\kappa$ is proportional to the conformal dimension of the corresponding boundary operator, $\beta$ is also proportional to the conformal dimension.  This relation can be understood qualitatively as follows. The boundary scale transformation $x_{\mu} \rightarrow a x_{\mu}$ has the same effect as the scale transformation along the $z$-direction $z \rightarrow z/a$, because of the isometry under the bulk scale transformation $x_{\mu} \rightarrow a x_{\mu}, z \rightarrow a z$. Also, near the boundary, the tunneling geodesic is parallel to the $z$-direction and takes the form $z=2 \alpha \exp{(\beta \tau)}$. Thus, a scale transformation of $z$ corresponds to a translation of the worldsheet parameter $\tau$. Then, because $\beta$ is the generator of the $\tau$ translation, the conformal dimension of the boundary operator and $\beta$ are directly related near the boundary.

\section{Two-point functions of the boundary operators corresponding to spinning strings}
In this section, we study the relation between the worldsheet action of the tunneling spinning string solution and the two-point function of the boundary operator. If we use Dirichlet boundary conditions in the string theory case, the boundary operator will be the Wilson loop from the holographic point of view. These two values should be directly connected by the relation
\begin{equation}
\langle W(C_1)W(C_2) \rangle_{\textrm{conn}} \sim e^{-S} \label{rer},
\end{equation}
where $S$ is the worldsheet action of this solution and $C_1$ and $C_2$, which form its boundaries, are loops
corresponding to string configurations.\footnote{The Gross-Ooguri transition \cite{9805129} occurs when the boundary objects are extended in the AdS direction.
However, in our work, we treat objects that are extended in $S^5$ space but are point-like in the AdS direction. 
Thus, the Gross-Ooguri transition does not occur, but a semiclassical calculation is still valid, because it is extended in $S^5$ space.} This is, of course, a
natural extension of the ansatz proposed in Refs. 6) and 7). From this point of view, double Wilson loop correlators have already been calculated for some special cases \cite{9809188,9904149}. We apply this relation to the spinning string/spin chain correspondence. In the spinning string case, the solution has fixed angular momenta in the $S^5$ directions, and therefore the boundary conditions are not Dirichlet conditions for some directions. For this reason, the boundary operator should not be the usual Wilson loop but, instead, an operator which has fixed R-charge. This operator should be related to the spin chain.

\subsection{Double Wilson loop correlator and OPE}
Before calculating the worldsheet action, we discuss the double Wilson loop correlator within the field theory description. We assume the relation between the usual Wilson loop and the fixed R-charge operator in order to justify the application of the holographic relation (\ref{rer}) to the spinnig string case.

When we consider small Wilson loops, we can expand them as \cite{9809188}
\begin{equation}
W(C)=\langle W(C) \rangle \left[1+\sum_{i, n}c_i^{(n)}a^{\Delta_i^{(n)}}\mathcal{O}_i^{(n)}(x)\right], 
\end{equation}
where $a$ is the characteristic size of the Wilson loop, $\{ \mathcal{O}_i^{(n)} \}$ is a set of operators with conformal weights $\Delta_i^{(n)}$, and $c_i^{(n)}$ are the coefficients, which depend on each loop $C$. Then, the connected part of the correlator of the double Wilson loops becomes
\begin{equation}
\frac{\langle W(C_1)W(C_2) \rangle_{\textrm{conn}}}{\langle W(C_1)\rangle \langle W(C_2) \rangle}=\sum_{i;l, m}c_i^{(m)}c_i^{(n)}a^{\Delta_i^{(m)}+\Delta_i^{(n)}}\langle \mathcal{O}_i^{(m)}(\alpha)\mathcal{O}_i^{(n)}(-\alpha) \rangle.  \label{ope}
\end{equation}
Here, $2 \alpha$ is the distance between the two loops. Thus, we can write the Wilson loop operator as an infinite sum of the two-point functions of the operators which have various definite conformal dimensions. Note that the scale parameter $a$ appears here because of the dimensionless property of Wilson loops.

In general, we have to choose some special configurations for the loops $C_1$ and $C_2$. Here, we want to consider the particular case in which the boundary objects have the definite R-charge $(J_1,J_2)$. Actually, these are not Wilson loops in the naive sense, because we fix the angular momenta with respect to $S^5$, not the loops $C_1$ and $ C_2$. More precisely, we are considering an object like
\begin{equation}
\mathcal{O}_b(J_i)=\int [d \mathcal{C}] \Psi_{J_i}(\mathcal{C}) W(\mathcal{C}), \label{gwil}
\end{equation}
where $\Psi_{J_i}$ is the wave function corresponding to a state of definite angular momenta, and the integration should be performed over the loops extending in the directions that are related to the scalar fields of the gauge theory. 
The normalization factor $\langle W \rangle$ can be included in the wave function $\Psi_{J_i}$. The choice of the wave function $\Psi_{J_i}$ is regarded as the choice of the boundary condition for the classical string solution near the boundary $z = 0$, which corresponds to a particular gauge invariant operator $\mathcal{O}_i^{(n)}$ with definite conformal dimension. It is desirable to develop a systematic method to explicitly construct the wave functions $\Psi_{J_i}(\mathcal{C})$ corresponding to spinning strings, but this is not necessary for the purpose of the present work.

\subsection{Calculation of the worldsheet action}
Now, we compute the worldsheet action of the tunneling spinning string solution. 
This action is given by
\begin{equation}
S=\frac{\sqrt{\lambda}}{2}\int_{-\infty}^{\infty}d \tau \int_0^{2 \pi}\frac{d \sigma}{2 \pi}\left( 
\eta(\frac{\dot{z}^2+\dot{t}^2}{z^2}-\dot{\varphi_1}^2\cos^2{\psi}-\dot{\varphi_2}^2\sin^2{\psi})+\frac{1}{\eta}{\psi '}^2\right). 
\end{equation}
The two fixed angular momenta are
\begin{equation}
J_1=-\sqrt{\lambda} \int \frac{d \sigma}{2 \pi}\eta \dot{\varphi_1} \cos^2{\psi}, J_2=-\sqrt{\lambda} \int \frac{d \sigma}{2\pi}\eta \dot{\varphi_2} \sin^2{\psi}. 
\end{equation}
As in Ref. 32), with the condition of two fixed angular momenta at the boundary, we have to perform a Legendre transformation of the coordinates $\varphi_1$ and $\varphi_2$. This yields 
\begin{eqnarray}
\bar{S}&=&S-\int d \tau (J_1 \dot{\varphi_1}+J_2 \dot{\varphi_2})\\
&=&\frac{\sqrt{\lambda}}{2} \int d \tau \frac{d \sigma}{2 \pi}  \nonumber \\
&& \times \left(\eta \left( \frac{\dot{z}^2+\dot{t}^2}{z^2} \right)
+\frac{1}{\eta}\left({\psi '}^2+\frac{J_1^2\cos^2{\psi}}{\lambda(\int \frac{d \sigma}{2\pi}\cos^2{\psi})^2}+\frac{J_2^2\sin^2{\psi}}{\lambda(\int \frac{d \sigma}{2\pi}\sin^2{\psi})^2}\right)
\right). 
\end{eqnarray}
Solving the equation of motion for $\eta$, we obtain
\begin{equation}
\eta=\sqrt{\frac{({\psi '}^2+\frac{J_1^2 \cos^2{\psi}}{\lambda(\int \frac{d \sigma}{2\pi}\cos^2{\psi})^2}+\frac{J_2^2 \sin^2{\psi}}{\lambda(\int \frac{d \sigma}{2\pi}\sin^2{\psi})^2})}{\frac{\dot{z}^2+\dot{t}^2}{z^2}}}. 
\end{equation}
Note that $\eta$ is independent of $\tau$ and $\sigma$.

From the above constraint, we obtain the action in the form 
\begin{equation}
\bar{S}=\sqrt{\lambda}\int_{-\infty}^{\infty} d \tau \eta \beta^2. \label{sol}
\end{equation}
Hence, this action is infinite, because of the infinite length of the AdS trajectory. 
Therefore, we must regularize this expression in order to obtain the dependence of the distance between the two loops. 

\subsection{Regularization}

In the case of a one-point correlator \cite{9803001, 9803002}, the physical value is obtained by subtracting the infinite quark mass contribution. In our case, we have to subtract similar infinite contributions.

\begin{figure}
\centering
\includegraphics[bb=0 0 270 280,clip]{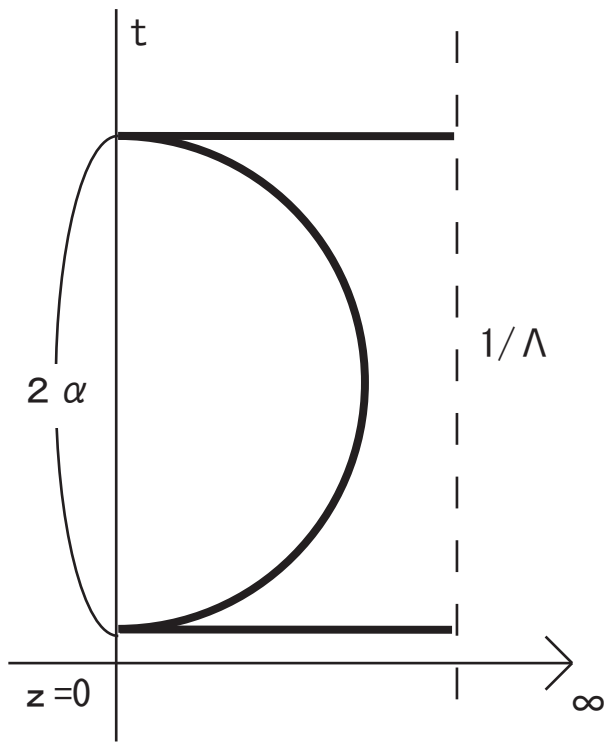}
\caption{AdS trajectories of the two-point function and one-point functions \label{fig:trajectory}}

\end{figure}

In Eq. (\ref{tuneom}), we can also consider a solution with constant $t$,
\begin{equation}
z=\frac{1}{\Lambda}\exp{(\pm \tilde{\beta} \tilde{\tau})},\hspace{12pt} t= \textrm{const},
\end{equation}
where $\Lambda$ is an arbitrary parameter. Because the $S^5$ part is unchanged, 
we have to choose $\tilde{\beta}=\beta$ from the constraints. We can use
this solution in order to subtract infinities. 
Explicitly, we consider the ratio of the two-point function determined by (\ref{sol}) to the product of the two one-point functions whose boundaries are at $t=
\pm \alpha$ (Fig. \ref{fig:trajectory}), corresponding to the two ends at the
boundary (see Fig. 1).

Here, we choose the range of $\tilde{\tau}$ as $-\infty \le \tilde{\tau} \le 0$ for $t=-\alpha$. In order to connect to the boundary, we must choose the $z=\frac{1}{\Lambda}e^{\beta \tilde{\tau}}$ solution. Similarly, for $t=\alpha$, we choose $0 \le \tilde{\tau} \le \infty$ and the $z=\frac{1}{\Lambda}e^{-\beta \tilde{\tau}}$ solution. Then, we can interpret $\Lambda$ as the infrared cutoff for one-point functions.

If we define $S_{1(t)}$ as the action for the one-point function at $t$, the regularized action becomes
\begin{eqnarray}
\bar{S}_{\textrm{reg}}&=&\bar{S}-(\bar{S}_{1(-\alpha)}+\bar{S}_{1(\alpha)}) \nonumber\\
&=&\sqrt{\lambda} \eta \left(\lim_{T \rightarrow \infty}\int_{-T}^{T}-\lim_{\tilde{T} \rightarrow \infty}(\int_{-\tilde{T}}^0+\int_0^{\tilde{T}})\right)\beta^2 \label{sreg}. 
\end{eqnarray}
To relate $T$ and $\tilde{T}$, we assume that $z$ at $\tau=T$ for the original solution is the same as $z$ at $\tilde{\tau}=\tilde{T}$ for the regulator solutions:
\begin{eqnarray}
\frac{1}{\Lambda}\exp{(-\beta \tilde{T})}&=&\frac{\alpha}{\cosh{\beta T}} \nonumber\\
& \rightarrow& 2 \alpha \exp{(- \beta T)}  {(T \rightarrow \infty)}.
\end{eqnarray}
Solving this equation, we obtain
\begin{equation}
\tilde{T}=T-\frac{1}{\beta}\log{(\Lambda (2\alpha))}.
\end{equation}
Substituting this into (\ref{sreg}), we get
\begin{eqnarray}
\bar{S}_{reg}&=&\frac{2\sqrt{\lambda}\eta}{\beta}\log{(\Lambda (2 \alpha))} \times \beta^2 \nonumber\\
&=&2 \sqrt{\lambda} \eta \beta \log{(\Lambda (2 \alpha))}. \label{reg}
\end{eqnarray}

\subsection{Two-point function in the bulk description}
In the AdS/CFT context, we can consider
\begin{equation}
\langle \mathcal{O}_b(- \alpha;J_i) \mathcal{O}_b(\alpha;J_i)\rangle = \exp{(-\bar{S}_{\textrm{reg}})}, \label{douwil}
\end{equation}
where $-\alpha$ and $\alpha$ are the positions of the loops.

From the result (\ref{reg}), we obtain
\begin{equation}
\exp{(-\bar{S}_{\textrm{reg}})}=
{\left(\frac{1}{2 \alpha \Lambda}\right)}^{2 \sqrt{\lambda} \eta \beta}. \label{dim}
\end{equation}
Substituting this result into (\ref{douwil}), we get
\begin{equation}
\langle \mathcal{O}_b(-\alpha;J_i)\mathcal{O}_b(\alpha;J_i) \rangle = {\left(\frac{1}{2 \alpha \Lambda}\right)}^{2 \sqrt{\lambda} \eta \beta}. 
\label{last}
\end{equation}
This implies that $\mathcal{O}_b(\pm \alpha;J_i)$ are simply local operators corresponding to $SU(2)$ spin chain states with the definite conformal dimensions $\Delta = \sqrt{\lambda}\eta \beta$. The quantity $\beta$ is determined by $J_1$ and $J_2$ and the string configuration, as in $\S$2.1. If we choose a string configuration, for example, a folded string or a circular string, we can find the corresponding Bethe roots configuration of the $SU(2)$ spin chain state. This is the same as usual spinning string/spin chain correspondence. Information concerning the holographic configuration mapping between string/spin chain coherent states may be contained in the wave equation (\ref{gwil}), but its explicit expression is not necessary to get the result (\ref{last}).

\section{Conclusion}
In this work, we have calculated the two-point function of the boundary operator 
using the holographic method for correlation functions, where the boundary operators are in states of the spin chain with definite conformal dimensions. Because the spinning string/spin chain correspondence has not previously been considered from the viewpoint of correlation functions, we believe that the present work is useful for clarifying this aspect of the problem.

As future works, there are some possible applications of the method employed here. In Refs. 25)-28), three-point functions of pp-wave string theory are studied by using the holographic relation. We may be able to investigate the three-point functions of spin chains in a similar manner. 
Also, we should investigate quantum effects by taking into account fluctuations from tunneling geodesics from boundary to boundary. Since
in the bulk picture the Hamiltonian along a geodesic is not identical to the dilatation operator, except near the boundary, these effects may affect the two-point function. It would be interesting to study whether this resolves the three-loop discrepancy. We may also study the holographic two-point function in cases other than AdS space by considering an arbitrary D$p$-brane background, as done in Ref. 32) in the pp-wave limit. 

\section*{Acknowledgements}
I would like to thank T. Yoneya for enlightening discussions and helpful
advice for improving the manuscript. I also thank Y. Mitsuka, A. Miwa, Y. Ohse and H. Shimada for useful comments. The present work is supported in part by JSPS Research Fellowship for Young Scientists. 

%

\end{document}